\documentclass[conference, a4paper]{IEEEtran}
\IEEEoverridecommandlockouts
\usepackage{cite}
\usepackage{amsmath,amssymb,amsfonts}
\usepackage{algorithmic}
\usepackage{booktabs}
\usepackage{graphicx}
\usepackage{textcomp}
\usepackage{xcolor}
\usepackage{array}
\usepackage{subcaption}
\usepackage{enumitem} 
\usepackage{cleveref}

\usepackage{geometry}
\def\BibTeX{{\rm B\kern-.05em{\sc i\kern-.025em b}\kern-.08em
    T\kern-.1667em\lower.7ex\hbox{E}\kern-.125emX}}
\begin{document}

\title{Benchmark Static API Call Datasets for Malware Family Classification\\
}

\makeatletter
\newcommand{\linebreakand}{%
 \end{@IEEEauthorhalign}
  \hfill\mbox{}\par
  \mbox{}\hfill\begin{@IEEEauthorhalign}
  
}
\makeatother
\newcommand\correspondingauthor{\thanks{{*}Corresponding author: 20151709009@stu.khas.edu.tr}}

\author{
  \IEEEauthorblockN{1\textsuperscript{st} Buket Gen\c{c}ayd{\i}n}
  \IEEEauthorblockA{\textit{Computer Engineering} \\
    \textit{Gebze Technical University}\\
    Kocaeli, Turkey \\
    b.gencaydin2019@gtu.edu.tr}
  \and
  \IEEEauthorblockN{2\textsuperscript{nd} Ceyda Nur Kahya}
  \IEEEauthorblockA{\textit{Management Information Systems} \\
    \textit{Kadir Has University}\\
    {\.I}stanbul, Turkey \\
    20161708007@stu.khas.edu.tr}
  \and
  \IEEEauthorblockN{3\textsuperscript{rd} Ferhat Demirk{\i}ran}
  \IEEEauthorblockA{\textit{Department of Cyber Security} \\
    \textit{Kadir Has University}\\
    {\.I}stanbul, Turkey  \\
    ferhat.demirkiran@khas.edu.tr}
  \linebreakand 
  \IEEEauthorblockN{4\textsuperscript{th} Berkant D\"{u}zg\"{u}n \correspondingauthor{*}}
  \IEEEauthorblockA{\textit{Management Information Systems} \\
    \textit{Kadir Has University}\\
    {\.I}stanbul, Turkey \\
   20151709009@stu.khas.edu.tr}
  \and
  \IEEEauthorblockN{5\textsuperscript{th} Aykut \c{C}ay{\i}r }
  \IEEEauthorblockA{\textit{Management Information Systems} \\
    \textit{Kadir Has University}\\
    {\.I}stanbul, Turkey \\
    aykut.cayir@khas.edu.tr}
  \and  
  \IEEEauthorblockN{6\textsuperscript{th} Hasan Da\u{g}}
  \IEEEauthorblockA{\textit{Management Information Systems} \\
    \textit{Kadir Has University}\\
    {\.I}stanbul, Turkey \\
    hasan.dag@khas.edu.tr}
     
}

\maketitle 

\begin{abstract}

Nowadays, malware and malware incidents are increasing daily, even with various antivirus systems and malware detection or classification methodologies. Machine learning techniques have been the main focus of the security experts to detect malware and determine their families. Many static, dynamic, and hybrid techniques have been presented for that purpose. In this study, the static analysis technique has been applied to malware samples to extract API calls, which is one of the most used features in machine/deep learning models as it represents the behavior of malware samples. 

Since the rapid increase and continuous evolution of malware affect the detection capacity of antivirus scanners, recent and updated datasets of malicious software became necessary to overcome this drawback.  
This paper introduces two new datasets: One with 14,616 samples obtained and compiled from VirusShare and one with 9,795 samples from VirusSample.
In addition, benchmark results based on static API calls of malware samples are presented using several machine and deep learning models on these datasets.
We believe that these two datasets and benchmark results enable researchers to test and validate their methods and approaches in this field.


\end{abstract}


\section{Introduction}

Malware is the short term used for malicious software, also known as a harmful code intentionally developed to harm a computing system.
Malware attacks can gain access to the system, disrupt system services, deny service, steal confidential information, and destroy resources, with irreversible consequences \cite{jang2014survey,aslan2020comprehensive}. 
To better understand how malware might infiltrate devices, computers, or systems, malware families must be known as family members shows similar behavior. Security researchers and incident responders can take preventative measures against malware by being able to identify its family, which could make the entire analysis process more straightforward \cite{han2019maldae}.

With the improvement of antivirus scanners to detect malware samples and classify their families, malware authors have started to change the nature of malware samples to evade these detection/classification mechanisms by using several techniques such as code obfuscation, polymorphism, and metamorphism. Thus, the need for new malware datasets is increasing as the complexity of the novel malware family samples grows correspondingly \cite{ucci2019survey}.

The three types of analyses used to extract the malware features are static, dynamic, and hybrid. Static technique analysis the structure of the malware to determine its behavior without running it, whereas the dynamic technique analysis the behavior of the malware during its execution, generally using a sandbox approach. On the other hand, hybrid techniques combine static and dynamic analysis aspects \cite{sihwail2018survey}. One feature extracted from malware is API calls, which represent each malware sample's behavioral pattern. Thus, these API calls are one of the most-frequently-used attributes with machine learning techniques in malware detection and classification.

Major contributions of this paper are:
\begin{itemize}
\item Presenting two alternative datasets \footnote{https://github.com/khas-ccip/api\_sequences\_malware\_datasets} that researchers can use in their studies on malware family classification based on API calls with increased current and diversified malware samples.
\item To our best knowledge, the dataset obtained from VirusSamples contains the most up-to-date malware samples based on API calls.
\item Proposing a new way to use cloud platforms to overcome the daily API key limit of VirusTotal without overloading or abusing the system to build a new dataset more efficiently.
\item Presentation of benchmark classification results with machine and deep learning techniques.

\end{itemize}
The rest of the paper is as follows: The related work and background of other research are discussed in Section~\ref{section:rw}. Section~\ref{section:m} presents features of the new datasets and the tools and steps for data construction and also  Google Cloud Platform for multi-user to increase efficiency in malware data collection. At Section~\ref{section:models} the models for benchmarking are introduced and the result are shown. Finally, Section~\ref{section:conclusion} concludes this paper.

\section{Related Work}
\label{section:rw}

In the malware detection operation, understanding malware behavior is one of the vital parts. The usage of API calls obtained with dynamic analysis is one of the most used approaches in malware analysis since they represent malware behavior \cite{b1}. \\
API Call sequences obtained by tracing the sequences of calls are used for generating specific behavioral patterns for each file. The sequences of API Calls collected through the dynamic approaches are used with data mining, machine learning, or deep learning techniques to detect or classify malware \cite{b2}. \\ 
The techniques mentioned above need to be applied on datasets created by API Call sequences. \\ 
In \cite{b3}, sequences of Windows Operating System API calls are obtained within the cuckoo sandbox isolated environment for each malware file. Malware family labels are determined using unique hash codes of each malware on the VirusTotal website. In total, there are 7,107 samples, which contain hash codes of malware, Windows operating system API Call sequences, and their malware family classes. \\  
In \cite{b4}, 42,797 malware API call sequences and 1,079 goodware API call sequences are obtained via cuckoo sandbox for dynamic malware analysis. Instead of using whole API call sequences, the first 100 non-consecutive API call sequences are extracted from the parent processes to reduce complexity and detect the malicious pattern as quickly as possible.
The generated dataset containing hashcodes, label (malware or goodware), and 100 API Calls for each sample was used for binary malware classification.

\section{Methodology}
\label{section:m}
\subsection{Datasets}
\label{section:ds}

In this study, we present two new datasets. These datasets contain hashcodes, API calls, and families of malware. Two different sites, VirusShare and VirusSample, are leveraged to extract MD5 hashcodes of malware samples. 
VirusShare is a malware repository of live malicious codes; all the samples are in ZIP format, with password protection. VirusSample generated malware samples robustly with various collection methods by processing more than 150,000 malware daily. For both sites to access the malware samples, a request must be sent from within sites; after the approval, the malware hashes are provided.
VirusShare and VirusSample datasets are constructed with MD5 hashes of malware, 14,616 samples collected from VirusShare and 9,795 samples collected from VirusSample sites, respectively; hence the datasets are named as VirusShare and VirusSample. 
VirusTotal site is used to determine the malware families by MD5 hashcode of malware samples. Table \ref{table:classDistribution} shows the total distribution of malware families for VirusShare and VirusSample datasets.

\begin{table}[ht]
\centering
\caption{Distribution of malware families.}
\begin{tabular}{@{}ccc@{}}
\toprule
\multicolumn{1}{c}{\textbf{Malware Family}} & \textbf{VirusShare } & \textbf{VirusSample } \\ \midrule
Trojan                                    	& 8,919		         & 6,153			     \\
Virus                                        & 2,490        & 2,367				   	       \\
Adware                                     & 908         & 222			      \\
Undefined                        &	577	         & N/A		   	       \\
Worms                                 & 524         & 441 \\
Backdoor                                       & 510	        & 447		                    \\
Downloader 		          & 218	          & 31			     \\
Agent                                    & 	165	         & 102		       \\ 
Ransomware                        &	115	         & 10		   	       \\
Riskware                        &	85	         & 4		   	       \\
Spyware                        &	45	         & 11		   	       \\
Dropper                        &	40	         & 4		   	       \\
Crypt                        &	10	         & 2		   	       \\
Keylogger                        &	7	         & 1		   	       \\
Rootkit                        &	3	         & N/A		   	       \\
\midrule
\textbf{Total}               & \textbf{14,616}                        & \textbf{9,795}              
\end{tabular}
\label{table:classDistribution}
\end{table}


VirusTotal is a service that analyzes suspicious files and facilitates real-time detection of malware content to detect malware families using over 70 antivirus scanners. The public API key of VirusTotal is a free service, available for any website or application but limited to 500 requests per day with a rate of 4 requests per minute.

The API calls of the malware samples are obtained using the Python module PEfile, which extracts API calls from a program's Portable Executable (PE) file header; hence, these API calls are not in execution call sequence order. Since these operations are one of the static analysis techniques, the obtained API calls are static API call sequences. 

Example malware samples in the VirusShare dataset are shown in Figure \ref{fig:figure1} to gain a general insight into datasets.

\begin{figure}[h]
\centering
\includegraphics[width=0.50\textwidth]{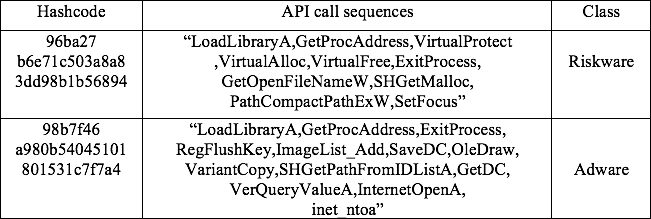}
\caption{Two malware sample examples from VirusShare dataset.}
\label{fig:figure1}
\end{figure}

\subsection{Dataset Construction}

The dataset construction steps are described below.

\begin{enumerate}[label=(\roman*)]
\item In each ZIP file obtained from VirusShare and VirusSample, malware samples are represented with their MD5 hashcodes. 
 The MD5 hashcodes of the malware samples are written to a text file in groups of 500 as the public API key of VirusTotal is limited to 500 requests per day. 

\item 
  Malware families are found by processing MD5 hashcodes in the VirusTotal system. VirusTotal analyzes these MD5 hashcodes to detect
malware families using over 70 antivirus scanners. The result of each antivirus scanner is checked, and the most frequently repeated family is marked as the malware family (Figure \ref{maininf}). The response of an antivirus scanner might be "undefined".

\item After identifying which family the malware belongs to via MD5 hashcodes, the malware's API calls are extracted using the Python module PEfile.

\item  Datasets with three sections; the MD5 hashcodes of malware samples, API calls from PEFile module in Python, and the malware family from VirusTotal, are gathered in CSV format.

\end{enumerate}


\begin{figure}[h]
\centering
\includegraphics[width=0.50\textwidth]{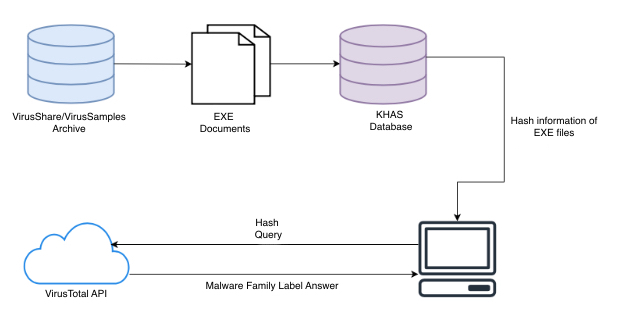}
\caption{Data construction infrastructure.}
\label{maininf}
\end{figure}

\subsection{Google Cloud Platform for Multi-User Malware Data Collection}
\label{section:google}
The VirusTotal system has a daily 3 API key limit which allows analyzing 1,500 MD5 hashcodes at most. Therefore we found a new approach to overcome this limitation.
 The steps for data construction that run on virtual machines moved into the Google Cloud Platform to increase the daily number of analyses.
The three main benefits of using the Google Cloud Platform are listed below:
\begin{itemize}
  \item The system on the cloud is an exact copy of the machine running on the virtual with better computing power and capacity.
  \item There is no IP conflict in the platform, which provides an increasing number of analyses by opening multiple virtual environments.
  \item The cloud allows downloading and analyzing large files that belong to different years \cite {mwcloud}.
\end{itemize}

A firewall rule must be established to allow downloading large files in cloud systems before launching the virtual machine. After that, the zipped system can be uploaded into the cloud platform to run the virtual machine. The image of the machine has to be taken after the installation to reduce the repeating steps by using the multiple machine feature of the cloud. Thus, the analysis continues without interruption while running 5 or 6 machines in the cloud system by using 3 API keys for each.

\begin{figure}[h]
\centering
\includegraphics[width=0.48\textwidth]{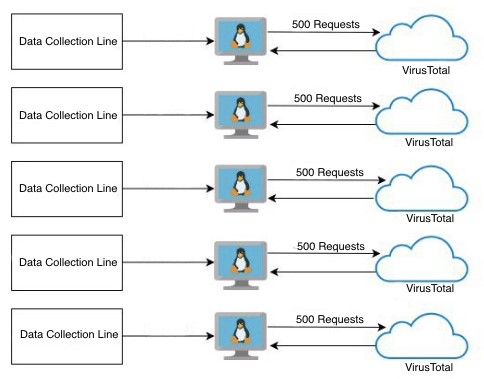}
\caption{Google Cloud Platform approach, for a single API key in 5 machines.}
\label{cloud}
\end{figure}

Figure \ref{cloud} shows the explained approach. This method assigns an IP address to each machine, thus preventing the conflict of IP addresses. Moreover, by using 3 API keys on each machine, 1,500 hashcodes can be analyzed in the VirusTotal system.

\section{Experiment and Results}
\label{section:models}
For benchmarking VirusShare and VirusSample datasets, simple preprocessing is applied. In both datasets, malware families with less than 100 repetitions and samples marked as undefined by VirusTotal are removed (Table \ref{table:classDistribution100}).

In addition, since the distributions of malware families are imbalanced, balanced versions of the two datasets are constructed by placing an upper limit of 300 for the malware families (Table \ref{table:classDistribution300}).

The benchmark results of four datasets are presented as balanced versions of these datasets constructed in addition to original imbalanced versions.
\begin{table}[ht]
\centering
\caption{Distribution of Malware Families After Preprocessing.}
\begin{tabular}{@{}ccc@{}}
\toprule
\multicolumn{1}{c}{\textbf{Malware Family}} & \textbf{VirusShare} & \textbf{VirusSample} \\ \midrule
Trojan                                    	& 8,919		         & 6,153	\\
Virus                                        & 2,490        & 2,367			\\
Adware                                     & 908         & 222			    \\
Worms                                 & 524         & 441 \\
Backdoor                                       & 510	        & 447		\\
Downloader 		          & 218	          & N/A			     \\
Agent                                    & 	165	         & 102		       \\ 
Ransomware                        &	115	         & N/A		   	       \\

\midrule
\textbf{Total}               & \textbf{13,849}                        & \textbf{9,732}              
\end{tabular}
\label{table:classDistribution100}
\end{table}

\begin{table}[ht]
\centering
\caption{Distribution of Malware Families in The Balanced Versions.}
\begin{tabular}{@{}ccc@{}}
\toprule
\multicolumn{1}{c}{\textbf{Malware Family}} & \textbf{VirusShare} & \textbf{VirusSample} \\ \midrule
Trojan                                    	& 300		         & 300		\\
Virus                                        & 300	        & 300				\\
Adware                                     & 300	         & 222			    \\
Worms                                 & 300	         & 300	 \\
Backdoor                                       & 300		        & 300			\\
Downloader 		          & 218	          & N/A			     \\
Agent                                    & 	165	         & 102		       \\ 
Ransomware                        &	115	         & N/A		   	       \\

\midrule
\textbf{Total}               & \textbf{1,998}                        & \textbf{1,524}              
\end{tabular}
\label{table:classDistribution300}
\end{table}
\subsection{Models For Benchmarking}

Seven different models are used for benchmarking. Two of these seven models are the traditional machine learning algorithms, Random Forest (RF) and Support Vector Machine (SVM). Another two are the ensemble models, Extreme Gradient Boosting (XGBoost) and Histogram-Based Gradient Boosting (HGBoost). Moreover, Long short-term memory (LSTM), a recurrent neural network using special units. Furthermore, two pre-trained transformer models, Bidirectional Encoder Representations from Transformers (BERT) and Character Architecture with No tokenization In Neural Encoder (CANINE). 

\subsubsection{Ensemble Models}
Ensemble learning offers a systematic solution to combine the predictive power of multiple learners. The resultant is a single model which gives the aggregated output from several models.

XGBoost and HGBoost are decision-tree-based ensemble machine learning algorithms that use a gradient boosting framework. Gradient boosting classification trees are becoming popular machine learning algorithms due to their ability in a wide range of implementation domains and easier management of model complexity by using tree depth and the number of trees \cite{gradboostcite,histgradcite}.

\subsubsection{Recurrent Neural Networks}

Recurrent neural networks (RNNs) are the widely used base architecture in natural language processing (NLP) as the information is retained through the previous step as input to the current step, unlike in traditional neural networks where the outputs and inputs are independent of each other. 


Since LSTM provides deeper processing of hidden states through specific units called memory cells, in contrast to the traditional RNNs, it is able to learn relatively long-term dependencies \cite{hochreiter1997long}.

\subsubsection{Pre-Trained Transformer Models}

Pre-trained Transformer Models (PTMs) are up-to-date deep learning models previously trained on the vast unlabeled text data to learn good universal representations. These PTMs are used to fine-tune the model on the downstream tasks with the learned representations \cite{ganesh2021compressing,qiu2020pre}.

\subsection{Experiment}

The four datasets presented in this study are divided into two parts, training and testing. 10\% of the datasets are allocated for testing. In the training phase, API call sequences are fed into the models to predict the class which is malware family (Figure \ref{fig:figure1}).
One of the essential steps on highly imbalanced datasets is preserving class distribution. Therefore, the data splitting process is performed in a stratified way. Also, for each dataset, we have applied Stratified 5-Fold strategy on training data, and hence 20\% of training data is used for validation on each iteration.
In BERT and CANINE, Stratified 5-Fold strategy is not leveraged since the PTMs have large number of parameters. 

The benchmark model result for the imbalanced and balanced version of the presented dataset, VirusShare and VirusSample, is given in Table \ref{table:shareorj} and Table \ref{table:sampleorj}, the best model according to F1-score and AUC score shown in bold font for each dataset. The confusion matrices of these best models are given in \Cref{fig:cm1,fig:cm2,fig:cm3,fig:cm4}.

\begin{table}[ht]
\centering
\caption{VirusShare Dataset Benchmark.}
\begin{tabular}{@{}ccccc@{}}
\toprule
& \multicolumn{2}{c}{Original Version} & \multicolumn{2}{c}{Balanced Version}\\ \midrule
\multicolumn{1}{c}{\textbf{Model}} & \textbf{F1-score} & \textbf{AUC score} & \textbf{F1-score} & \textbf{AUC score} \\ \midrule
RF                    & 0.6020  & 0.9334   & 0.6609  & 0.9304                 \\
SVM       & \textbf{0.7343}    & 0.9226    & \textbf{0.7581}    & 0.9404                  \\
XGBoost     & 0.7178      & \textbf{0.9666}  & 0.7525      & \textbf{0.9577}             \\
HGBoost     & 0.6952      & 0.9582    & 0.7470      & 0.9447              \\ 

LSTM           & 0.7007      & 0.9359  & 0.7007      & 0.9359                  \\
BERT             & 0.7068      & 0.9432   & 0.7447      & 0.8843               \\
CANINE          & 0.6955     & 0.9261     & 0.7284     & 0.9045              \\
\midrule
\end{tabular}
\label{table:shareorj}
\end{table}

\begin{table}[ht]
\centering
\caption{VirusSample Dataset Benchmark.}
\begin{tabular}{@{}ccccc@{}}
\toprule
& \multicolumn{2}{c}{Original Version} & \multicolumn{2}{c}{Balanced Version}\\ \midrule
\multicolumn{1}{c}{\textbf{Model}} & \textbf{F1-score} & \textbf{AUC score} & \textbf{F1-score} & \textbf{AUC score} \\ \midrule
RF                  & 0.5510  & 0.8962   & 0.8391  & 0.9688                 \\
SVM      & 0.7331    & 0.9640      & 0.8975     & 0.9782              \\
XGBoost     & 0.7351      & \textbf{0.9816}    & 0.9031       & \textbf{0.9941}                \\
HGBoost      & 0.7315      & 0.9776     & 0.8802       & 0.9841                 \\ 

LSTM             & \textbf{0.7788}     & 0.9682   & 0.8400       & 0.9478                   \\
BERT         & 0.7253      & 0.9590   & 0.8948         & 0.9708                 \\
CANINE         & 0.7182     & 0.9621   & \textbf{0.9086}     & 0.9828               \\
\midrule
\end{tabular}
\label{table:sampleorj}
\end{table}

The result indicates that SVM outperformed the other models according to the F1-score for both versions of the VirusShare dataset, whereas XGBoost achieved the highest AUC score for each dataset. On the other hand, in the VirusSample dataset, LSTM and CANINE obtained the highest F1-score in imbalanced and balanced versions, respectively.

\begin{figure}[ht]
\begin{subfigure}{0.24\textwidth}
  \centering
  \includegraphics[width=\linewidth]{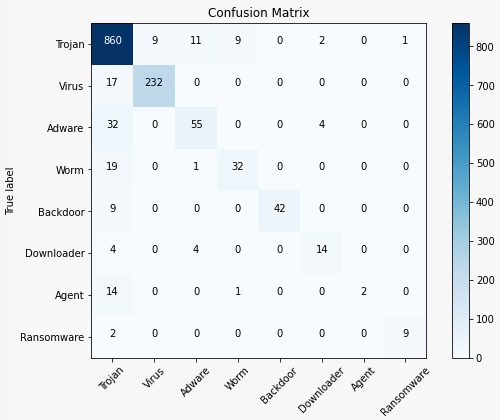}  
  \caption{SVM model}
\end{subfigure}
\begin{subfigure}{0.24\textwidth}
  \centering
  \includegraphics[width=\linewidth]{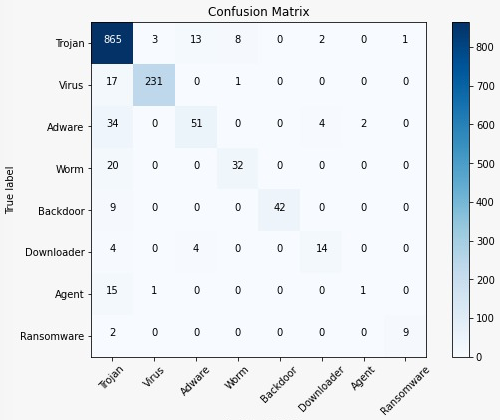}
  \caption{XGBoost model}
\end{subfigure}
\caption{Confusion matrix for the imbalanced version of VirusShare dataset, (a) the model achieving the highest F1-score, (b) the model obtaining the highest AUC score.}
\label{fig:cm1}
\end{figure}

\begin{figure}[ht]
\begin{subfigure}{0.24\textwidth}
  \centering
  \includegraphics[width=\linewidth]{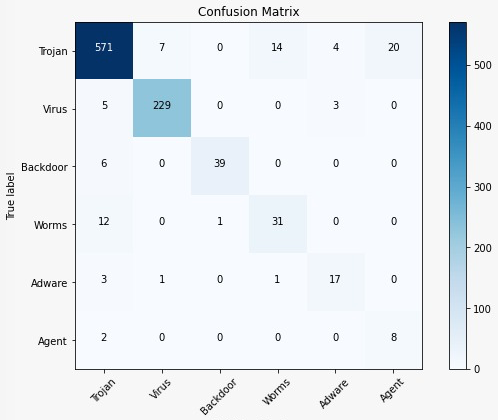}  
  \caption{LSTM model}
\end{subfigure}
\begin{subfigure}{0.24\textwidth}
  \centering
  \includegraphics[width=\linewidth]{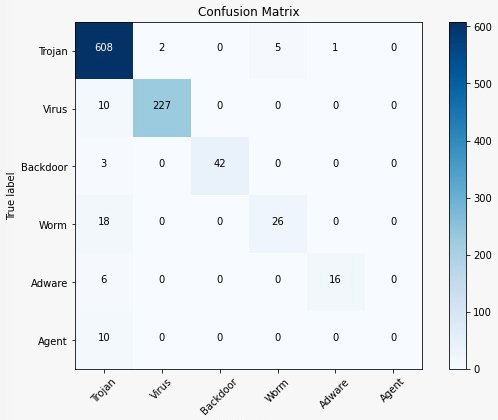}
  \caption{XGBoost model}
\end{subfigure}
\caption{Confusion matrix for the imbalanced version of VirusSample dataset, (a) the model achieving the highest F1-score, (b) the model obtaining the highest AUC score.}
\label{fig:cm2}
\end{figure}

\begin{figure}[!h]
\begin{subfigure}{0.24\textwidth}
  \centering
  \includegraphics[width=\linewidth]{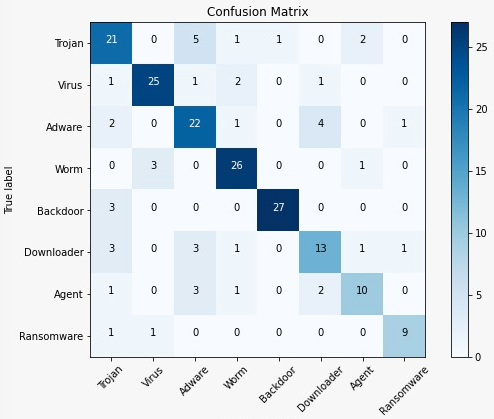}  
  \caption{SVM model}
\end{subfigure}
\begin{subfigure}{0.24\textwidth}
  \centering
  \includegraphics[width=\linewidth]{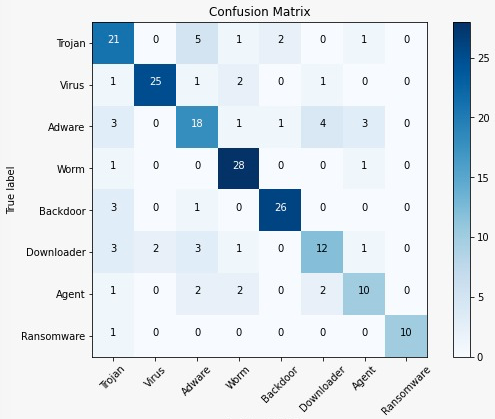}
  \caption{XGBoost model}
\end{subfigure}
\caption{Confusion matrix for the balanced version of \mbox{thatshouldnot}  dataset, (a) the model achieving the highest F1-score, (b) the model obtaining the highest AUC score.}
\label{fig:cm3}
\end{figure}

\begin{figure}[!h]
\begin{subfigure}{0.24\textwidth}
  \centering
  \includegraphics[width=\linewidth]{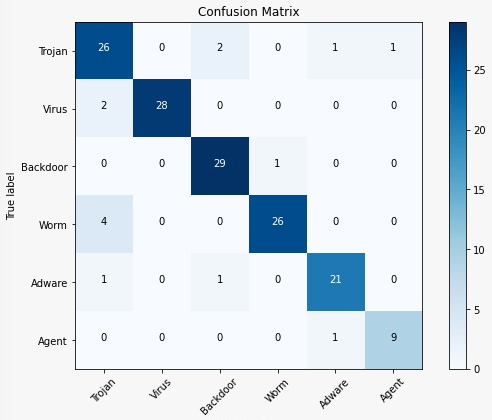}  
  \caption{CANINE model}
\end{subfigure}
\begin{subfigure}{0.24\textwidth}
  \centering
  \includegraphics[width=\linewidth]{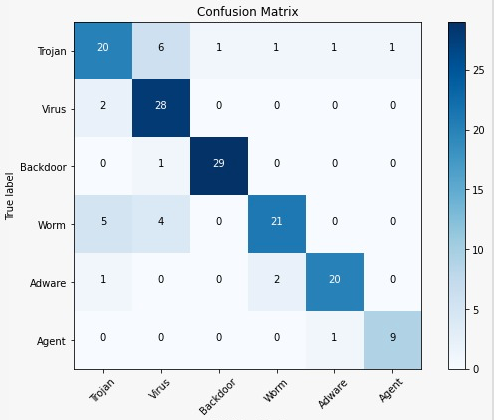}
  \caption{XGBoost model}
\end{subfigure}
\caption{Confusion matrix for the balanced version of VirusSample dataset, (a) the model achieving the highest F1-score, (b) the model obtaining the highest AUC score.}
\label{fig:cm4}
\end{figure}

\newpage
\section{Conclusion}
\label{section:conclusion}

This paper introduces two new static API call datasets, VirusShare and VirusSample, that consists of MD5 hashcodes, static API calls, and families of malware samples. A new approach is leveraged via the Google Cloud Platform, which allows us to launch multiple machines with different IP addresses and speed up the data gathering process. Therefore, VirusShare and VirusSample datasets are constructed with 14,616 and 9,795 malware samples, respectively. With that, benchmark results of multiple models from two versions, imbalanced and balanced, of the VirusShare and VirusSample datasets are presented.  

We sincerely hope that these two datasets will enable researchers to test their methods and compare the results in this paper.

\section*{Acknowledgment}
This work is supported by The Scientific and Technological Research Council of Turkey under the grant number 118E400.
 \bibliographystyle{elsarticle-num} 
 \bibliography{cite}

\vspace{12pt}

\end{document}